\documentclass[reprint,amsmath,amssymb,aps,pre,floatfix]{revtex4-2}

\usepackage[caption=false]{subfig}
\usepackage{graphicx}
\usepackage{amsmath}
\usepackage{amssymb}
\usepackage{braket}
\usepackage[table]{xcolor}
\usepackage{verbatim}

\begin{document}

	\title{Basis-free neural-network geminal and Jastrow factors for variational Monte Carlo}%

	\author{Jan Kessler}%
	\affiliation{Dynamics of Condensed Matter and Center for Sustainable Systems Design, Chair of Theoretical Chemistry, University of Paderborn, Warburger Str. 100, D-33098 Paderborn, Germany}
	\author{Thomas D. K\"uhne}
	\email{tkuehne@cp2k.org}
	\affiliation{Center for Advanced Systems Understanding (CASUS), Conrad-Schiedt-Straße 20, 02826 Görlitz}
	\affiliation{Helmholtz Zentrum Dresden-Rossendorf, Bautzner Landstraße 400, 01328 Dresden, Germany}
	\affiliation{Institute of Artificial Intelligence, Technische Universit\"at Dresden, Helmholtzstraße 10, 01069 Dresden, Germany}

	\begin{abstract}
	Neural-network quantum states offer a flexible route to compact many-electron wave functions, but their practical accuracy depends strongly on how fermionic antisymmetry, electron correlation, and optimization noise are treated.
	Here we combine an antisymmetrized geminal power (AGP) determinant with feed-forward neural networks that replace conventional basis-set expansions in the geminal and in two Jastrow-factor constructions.
	The resulting basis-free Jastrow--AGP ansatz is optimized by variational Monte Carlo and is designed to separate two tasks: the AGP part defines the nodal surface, while the neural-network Jastrow factor recovers dynamical correlation at fixed nodes.
	This separation makes it possible to distinguish errors associated with dynamical correlation from those caused by static, multireference correlation.
	Applications to the hydrogen molecule and the rectangular hydrogen tetramer show sub-millihartree accuracy when the AGP nodes are adequate, and expose the residual nodal limitation near the large-radius square geometry of the hydrogen tetramer.
	These results clarify where neural-network building blocks can improve a compact geminal ansatz and where additional nodal flexibility is required.
	\end{abstract}

	\maketitle

	\section{Introduction}
	The quantum mechanical description of realistic molecular systems by accurate electronic ground-state wave functions remains a central challenge in chemical physics. The difficulty arises not only from fermionic antisymmetry, but also from the need to describe dynamical and static electron correlation on an equal footing. Determinantal wave functions provide the required antisymmetry, yet a systematically improvable treatment of correlation often requires a rapidly growing expansion of determinants, leading to high-order polynomial or exponential scaling of the computational effort and making larger systems inaccessible.
	
	A particularly useful compromise is offered by compact wave-function forms that retain a single determinantal object while augmenting it with explicit correlation. A common choice is a Slater determinant \cite{slater_det} of one-particle orbitals, expanded in a finite basis set, multiplied by a Jastrow correlation factor \cite{jastrow_factor}. Here we instead use a determinant of two-particle functions, the antisymmetrized geminal power (AGP), originally introduced in Refs.~\onlinecite{pople_lj_hurley_agp} and later combined with a Jastrow factor in Ref.~\onlinecite{sorella_agp}. The resulting Jastrow--AGP (JAGP) wave function retains the computational simplicity of a single determinant while carrying a multiconfigurational character, and can be more accurate than a Jastrow--Slater ansatz for strongly correlated geometries \cite{genovese_h4-agp}.
	This construction is closely related to the resonating-valence-bond (RVB) picture, which originates in Pauling's chemical resonance concept and was later reformulated by Anderson for strongly correlated lattice systems \cite{pauling_rvb,Anderson1987RVB}. In that context, an RVB state is often represented as a Gutzwiller-projected BCS wave function, $\hat{P}_{G}\ket{\mathrm{BCS}}$, in which the BCS pairing state supplies singlet bonds and the Gutzwiller projector enforces strong local correlation \cite{Anderson2004PlainVanillaRVB}. In the present continuum molecular setting, the AGP determinant can be viewed as a number-projected BCS-like singlet-pairing state, while the Jastrow factor plays the role of a flexible positive correlation factor.
	This RVB/JAGP viewpoint has also been used in TurboRVB-based QMC studies, including our previous application to the ozone molecule and more recent work on correlated kagome materials, where replacing a Slater determinant by an AGP form explicitly exposes static correlation and pairing effects \cite{TurboRVB,TurboGenius,OzoneTurboRVB,Azadi2025Herbertsmithite}. Related QMC benchmarks on dense hydrogen further illustrate why compact, explicitly correlated many-body wave functions remain valuable beyond small molecular test cases \cite{Azadi2013Hydrogen,Azadi2020HydrogenEOS}.
	
	Computing observables from Jastrow-correlated wave functions entails the evaluation of high-dimensional integrals, which requires a scalable integration method for systems with many electrons. The standard choice is importance-sampled Monte Carlo (MC) integration \cite{metropolis_MC,hastings_MC}, a stochastic approach that is naturally suited to massively parallel execution. Variational Monte Carlo (VMC) combines this integration strategy with systematic optimization of trial-wave-function parameters according to the Rayleigh--Ritz variational principle \cite{mcmillan_VMC}; it forms one of the central branches of quantum Monte Carlo (QMC) methods \cite{ceperley_QMC,foulkes_QMC}, for which broad overviews and chemical applications are available in Refs.~\onlinecite{Needs2010QMC,Austin2012QMCReview}.
	A complementary stochastic route to correlated electronic structure is full configuration interaction quantum Monte Carlo (FCIQMC), introduced by Alavi and coworkers, which samples signed walker populations in Slater-determinant space and can reach finite-basis FCI-quality energies without storing the complete CI vector \cite{BoothThomAlavi2009FCIQMC}. Its initiator extension further reduced the walker populations required for large determinant spaces \cite{ClelandBoothAlavi2010FCIQMC}. Both determinant-space projector methods and fixed-node real-space QMC reflect the same underlying fermion-sign problem: amplitudes can often be improved efficiently, but the antisymmetric sign structure or, equivalently for real wave functions, the nodal hypersurface remains the central obstruction to systematically exact simulations. In contrast to that determinant-space projection philosophy, the present work asks how much of the same correlation problem can be captured by a compact, explicitly interpretable real-space VMC ansatz.
	
	The quality of a JAGP--VMC calculation still depends on the form of the Jastrow factor and on the basis used to represent the geminal. These choices affect both accuracy and cost, and optimal parameterizations are system dependent.
	In this work we explore artificial neural networks (ANNs), specifically feed-forward neural networks (FFNNs), as basis-free building blocks for both the Jastrow factor and the geminal determinant.
	This idea is motivated by the universal approximation properties of FFNNs \cite{hornik_approximation_1991} and by the rapid development of neural-network QMC simulations for continuous systems, including model systems, atoms, and molecules \cite{han_deepwf_2019,carleo_fermionic_2020,Pfau_FermiNet,kessler_2particle-nnvmc}, PauliNet extensions to excited states \cite{PauliNetExcitedStates}, transition-metal compounds \cite{PhysRevResearch.4.013021}, molecular excitations \cite{foulkes_better_ferminet_2020}, and periodic systems \cite{carleo_periodic_nnwf_2022,wilson_periodic_nnwf_2022_arxiv,li_periodic_nnwf_2022_arxiv}. While some of these architectures can reach a higher ultimate accuracy than the compact wave functions considered here, our goal is to assess whether neural-network components can improve a transparent JAGP form while keeping the nodal and correlation contributions physically interpretable.
	This places the present work between two active lines of development. On the one hand, highly expressive neural wave functions such as PauliNet and FermiNet seek near-black-box variational flexibility \cite{PauliNetExcitedStates,Pfau_FermiNet}. On the other hand, JAGP and TurboRVB-type wave-function forms retain a chemically interpretable pairing structure \cite{sorella_agp,TurboRVB,TurboGenius}. Our aim is closer to the latter: we trade some ultimate flexibility for a decomposition in which changes in the Jastrow factor can be interpreted separately from changes in the nodal surface carried by the AGP part.
	
	\section{Neural-network Jastrow--AGP ansatz and variational optimization}
	\label{sec:nnagp}
	For the wave functions in this work, we use the AGP determinant to provide the nodal surface and the required exchange antisymmetry:
	\begin{align}
	\label{eq:psi_jagp}
	\left|\Psi_{\mathrm{JAGP}}\right\rangle &= \Psi_{\mathrm{JF}}(1,\dots,N)\,\Psi_{\mathrm{AGP}}(1,\dots,N) \\
	&= \Psi_{\mathrm{JF}}\,
	\begin{vmatrix} \Psi_G(1, \frac{N}{2} + 1) & \dots  & \Psi_G(1, N) \\
		   \vdots    & \ddots & \vdots    \\
		\Psi_G(\frac{N}{2}, \frac{N}{2} + 1) & \dots  & \Psi_G(\frac{N}{2}, N)
	\end{vmatrix} \nonumber
	\end{align}
	Typically, the geminal function $\Psi_G$ is expanded in products of atomic basis functions. Here, by contrast, we employ a feed-forward neural network directly as the geminal. For a single hidden layer, such an FFNN can be written as
	\begin{align}
		\label{eq:dist_fed_nn}
		F_{NN}& (\mathbf{r}) = \\* 
		& a_{o} \left( \omega_{o,0} + \sum_{i=1}^{N_h} \omega_{o,i} a_{h,i} \left( \omega_{h, i, 0} + \sum_{j=1}^{N_d} \omega_{h,i,j} d_j(\mathbf{r}) \right) \right), \nonumber
	\end{align}
	where $d_j(\mathbf{r})$ are the $N_d$ electron--electron and electron--nucleus distances for the electronic coordinates $\mathbf{r}$, the $\omega$ are variational connection weights, $a(x)$ denotes the activation function, and $N_h$ is the number of hidden units. Since FFNNs possess universal approximation capabilities \cite{hornik_approximation_1991}, they can in principle approximate a broad class of geminal functions. They do not, however, automatically obey the pair-exchange symmetry required in an AGP. Because the geminal depends only on two particle positions, this symmetry can be imposed by forming either a sum,
	\begin{equation}
		\Psi_G(\mathbf{r}_1, \mathbf{r}_2) = F_{NN}(\mathbf{r}_1, \mathbf{r}_2) + F_{NN}(\mathbf{r}_2, \mathbf{r}_1)
	\end{equation}
	or a product,
	\begin{equation}
		\Psi_G(\mathbf{r}_1, \mathbf{r}_2) = F_{NN}(\mathbf{r}_1, \mathbf{r}_2)\,F_{NN}(\mathbf{r}_2, \mathbf{r}_1).
	\end{equation}
	Following preliminary tests, the product form was used for all results reported here. All FFNNs contained two hidden layers, with the number of hidden units varied as described below. We denote the neural-network-based $\Psi_{\mathrm{AGP}}$ by NNAGP.
	
	For the Jastrow part $\Psi_{JF}$ we consider two options, both again utilizing a FFNN. The first variant
	\begin{equation}
		\Psi_{\mathrm{JF},P}(\mathbf{r}) = \Psi_{\mathrm{cusp}}(\mathbf{r})\,\exp\left[\sum_{i<j} P_{NN}(\mathbf{r}_i, \mathbf{r}_j)\right]
	\end{equation}
	which we call the pairing Jastrow factor. Here $P_{NN}(\mathbf{r}_i,\mathbf{r}_j)$ is a symmetrized FFNN of the same type as the geminal, and $\Psi_{\mathrm{cusp}}(\mathbf{r})$ is a simple cusp-correction factor,
	\begin{equation}
		\Psi_{\mathrm{cusp}}(\mathbf{r}) = \exp\left[U_{ee}(\mathbf{r}) + U_{en}(\mathbf{r})\right],
	\end{equation}
	where
	\begin{equation}
		U_{ee}(\mathbf{r}) = \sum_{i<j} \frac{|\mathbf{r}_j - \mathbf{r}_i|}{2(1 + b_{ee} |\mathbf{r}_j - \mathbf{r}_i|)} 
	\end{equation}
	and
	\begin{equation}
		U_{en}(\mathbf{r}) = -\sum_i^{N_e} \sum_j^{N_n} \frac{1 - \exp(-b_{en} |\mathbf{R}_j - \mathbf{r}_i|)}{b_{en}},
	\end{equation}
	with nuclear coordinates $\mathbf{R}$ and variational parameters $b_{ee}$ and $b_{en}$. The factor $\Psi_{\mathrm{cusp}}$ helps satisfy the electron--electron and electron--nucleus cusp conditions. In combination with the NNAGP, this wave-function form mirrors the JAGP structure used in Ref.~\onlinecite{genovese_h4-agp}, but replaces the basis-set expansions in the geminal and pairing functions by single FFNNs. This makes it possible to compare directly with the conventional basis-set construction and to isolate the effect of the neural-network representation.
	
	Going beyond simple substitutions, we also consider an NN-based all-body Jastrow factor that inherits ideas from our previous work: 
	\begin{equation}
		\Psi_{\mathrm{JF},A}(\mathbf{r}) = \Psi_{\mathrm{cusp}}(\mathbf{r})\,J_{NN}(\mathbf{r}), 	
	\end{equation}
	where $J_{NN}(\mathbf{r})$ is again an FFNN of the form in Eq.~\eqref{eq:dist_fed_nn}. Unlike the geminal and pairing functions, it uses all electronic coordinates simultaneously. In principle, the flexibility of the FFNN allows this all-body NNJF to recover all correlation that is accessible without changing the nodal surface fixed by the determinantal part.
	As for the geminal, an unconstrained all-body FFNN is not strictly exchange symmetric. Following the reasoning of our previous work \cite{kessler_2particle-nnvmc}, we rely on the approximate exchange symmetry that emerges under energy minimization when the FFNN output is constrained to be strictly positive by the output activation function. In that case the Jastrow factor does not alter the nodes imposed by the AGP determinant.
	This division of labor between an antisymmetric nodal carrier and a flexible positive correlation factor is also connected to our earlier fermionic and antisymmetric shadow-wave-function work, where auxiliary degrees of freedom were used to incorporate high-order many-body correlations while making the nodal/sign structure explicit \cite{Calcavecchia2014ShadowSign,Calcavecchia2015ShadowSlater,Calcavecchia2018ASWF}.
	The same observation gives a useful theoretical diagnostic. For a real wave function, multiplying by a positive Jastrow factor can change amplitudes, local correlations, and cusp behavior, but it leaves the nodal hypersurface unchanged. Energy improvements obtained by enlarging the all-body NNJF at fixed NNAGP therefore measure correlation recovery within a fixed-node manifold, in the same sense in which fixed-node errors are separated from projection errors in diffusion Monte Carlo \cite{foulkes_QMC,Needs2010QMC}. By contrast, residual errors that remain insensitive to the NNJF size point to missing nodal flexibility and require changes in the antisymmetric part, for example a richer geminal, additional determinants, backflow-like coordinate transformations, or more expressive neural antisymmetrization.
	A separate formal issue is size consistency. The present ansatz is variational, but it is not automatically size consistent or size extensive in the quantum-chemical sense. For two noninteracting fragments $A$ and $B$, strict size consistency would require $\Psi_{AB}\rightarrow\Psi_A\Psi_B$ and $E_{AB}=E_A+E_B$ in the dissociation limit. In the AGP--Jastrow form this amounts to a block-separable geminal together with Jastrow terms whose interfragment contributions vanish. The pairing Jastrow can represent such a cluster decomposition by suppressing cross terms, whereas a single global all-body NNJF does not enforce it by construction. Thus, size consistency should be regarded here as a property to test and, if needed, impose architecturally in larger applications, rather than as a theorem of the present finite-system implementation.
	
	In the following we denote $\Psi_{\mathrm{JF},P}$ and $\Psi_{\mathrm{JF},A}$ as pairing-NNJF and all-body-NNJF, respectively, and refer to the full NNAGP--NNJF product as the neural-network wave function (NNWF).
	
	For applications, the NNWFs are used as trial wave functions $\Psi_T(\mathbf{r})$ for the ground state $\Psi_0(\mathbf{r})$ of the nonrelativistic, time-independent electronic Schr\"odinger equation,
	\begin{equation}
		\hat{H}\ket{\Psi_T(\mathbf{r})} = E_T\ket{\Psi_T(\mathbf{r})},
		\label{eq:schroedinger}
	\end{equation}
	for a given Hamiltonian $\hat{H}$. The variational parameters are optimized by minimizing $E_T$ toward the ground-state energy $E_0$. This approach is justified by the Rayleigh--Ritz variational principle,
	\begin{equation}
		\label{eq:rrvarp}
		E_0 \leq E_T = \frac{ \braket{\Psi_T|\hat{H}|\Psi_T} }{ \braket{\Psi_T|\Psi_T} }.
	\end{equation}
	
	For larger systems, evaluating $E_T$ and other observables requires numerical integration of high-dimensional integrals. We perform these integrations by importance-sampled MC sampling in the Metropolis--Hastings framework, which is well suited to massively parallel execution. The price of this favorable scaling is statistical noise in the sampled observables and gradients.
	
	Energy minimization is carried out by minimizing the regularized cost function
	\begin{equation}
		C = E_{T} + \frac{\lambda_r}{N_{\omega}} \sum_{i=1}^{N_{\omega}} \omega_i^2
	\end{equation}
	with respect to the $N_\omega$ parameters $\omega$ of the trial wave function, which are mainly neural-network weights. The $L_2$ regularization with small prefactor $\lambda_r$ stabilizes the optimization.
	
	To optimize thousands of parameters in the presence of stochastic error, we use the Adam algorithm \cite{kingma_adam_2014}. The optimized weights are obtained from the exponential moving average recommended by Kingma and Ba (Sec.~7.2 of Ref.~\onlinecite{kingma_adam_2014}), reducing the influence of sampling noise on the final parameters. We use an analytically derived expression for the energy gradient \cite{kessler_2particle-nnvmc}, evaluate the resulting integrals by MC sampling, and obtain the neural-network derivatives, e.g., $\partial_{\omega_i}F_{NN}$, $\partial_{\omega_i}P_{NN}$, and $\partial_{\omega_i}J_{NN}$, by reverse-mode differentiation.
	
	Because successive samples in the MC walk are correlated, statistical uncertainties were estimated with a blocking analysis \cite{blocking_orig}. We used the efficient implementation of Ref.~\onlinecite{blocking_jonsson}, which supports large per-thread sample counts with negligible overhead, provided that the number of samples is a power of two.
	
	\section{Computational details}
	\label{sec:compdet}
	\begin{figure}
		\includegraphics[width=\columnwidth]{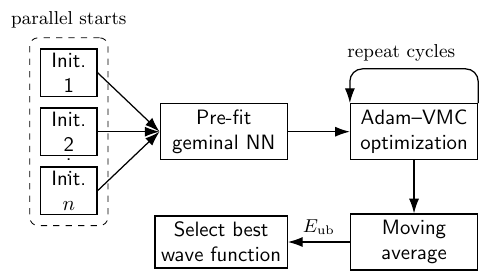}	
		\caption{Optimization protocol used for the neural-network wave functions. \label{fig:opt_illustration}}
	\end{figure}
	To assess the NNWF ansatz, we applied it to two hydrogen systems. The first is the hydrogen molecule, $H_2$, which is computationally inexpensive but still probes the transition from dynamical to static correlation along bond dissociation. The second is a rectangular arrangement of four hydrogen atoms, denoted $H_4$. Square $H_4$ was originally studied by QMC \cite{anderson_h4} and later established as a stringent wave-function benchmark \cite{gasperich_h4}. Recent data from a conventional JAGP ansatz \cite{genovese_h4-agp} and from FermiNet \cite{Pfau_FermiNet} make this system well suited to separate basis-representation errors, Jastrow flexibility, and nodal limitations.

	Both systems were treated within the Born--Oppenheimer approximation. In atomic units, the electronic Hamiltonian is
	\begin{align}
		\hat{H} =& - \frac{1}{2} \sum_{i=1}^{N_e} \nabla_{\mathbf{r}_i}^2
		- \sum_{i=1}^{N_e}\sum_{A=1}^{N_n} \frac{Z_A}{|\mathbf{r}_i-\mathbf{R}_A|} \\
		&+ \sum_{i<j}^{N_e}\frac{1}{|\mathbf{r}_i-\mathbf{r}_j|}
		+ \sum_{A<B}^{N_n}\frac{Z_A Z_B}{|\mathbf{R}_A-\mathbf{R}_B|},
		\nonumber
		\label{eq:molec_ham}
	\end{align}
	where $\mathbf{r}_i$ are electronic coordinates, $\mathbf{R}_A$ are fixed nuclear coordinates, and $Z_A$ are nuclear charges.
	
	For all calculations we used the optimization scheme shown in Fig.~\ref{fig:opt_illustration}. First, the geminal was preconditioned by a least-squares fit to a localized Gaussian form. This was followed by VMC energy minimization of the full wave function, as described in Sec.~\ref{sec:nnagp}, in cycles with a fixed number of Adam iterations. Each cycle was initialized from the final parameters of the preceding cycle. Several optimizations were run in parallel from different random initializations, and the wave function with the lowest upper energy confidence bound,
	\begin{equation}
		E_{\mathrm{ub}}=E_T+2\sigma_T,
	\end{equation}
	was selected for the final energy evaluation.
	
	The cyclic minimization consisted of 600 Adam cycles with 250 iterations each. The Adam parameters were $\alpha=0.0005$, $\beta_1=0.8$, $\beta_2=0.95$, and $\epsilon=10^{-8}$, using the notation of Ref.~\onlinecite{kingma_adam_2014}; the regularization factor was $\lambda_r=0.05$.
	
	The total number of MC sampling steps for energy and gradient evaluation in each Adam iteration was initially $40\times2^{10}$, distributed over 40 threads. This number was doubled after cycles 224, 374, 474, and 549, allowing fast early optimization and more accurate late-stage refinement. Final energy evaluations used $40\times2^{26}$ MC steps, i.e., approximately $2.7\times10^9$ steps, so that statistical error bars are smaller than the graphical resolution of the plots. The MC walkers used Gaussian all-particle moves; the step size was adjusted periodically to maintain an acceptance probability of approximately $50\%$. Energies and gradients were sampled every fourth step after burn-in periods of 500 steps per thread during optimization and 2000 steps per thread for final energy evaluation.
	
	All hidden units used the Softplus activation function \cite{dugas_softplus_2001}, $a(x)=\log(1+e^x)$. The output activation was exponential, except for the FFNNs in the pairing-NNJF, where a linear output was used.
	\section{Results and discussion}

	We organize the results so as to separate the representational accuracy of the neural-network components from the nodal limitations of the compact AGP carrier. In this discussion, dynamical correlation denotes the short-time avoidance of electrons within an otherwise qualitatively well-defined bonding pattern, including dispersion-like interactions between weakly interacting fragments. Static correlation, in contrast, denotes the near-degeneracy of several bonding patterns or determinants, as encountered when covalent bonds are stretched and a single-reference description loses its qualitative validity \cite{Calcavecchia2015ShadowSlater}. The $H_2$ dissociation curve first provides a two-electron validation problem for which the exact nodal structure is simple and for which neural-network VMC has already been shown to be highly accurate \cite{kessler_2particle-nnvmc}. The rectangular $H_4$ benchmarks then probe the same NNAGP--NNJF construction in a genuinely multielectron setting where intramolecular H--H stretching and intermolecular H$_2$--H$_2$ separation can be used to tune dynamical and static correlation separately \cite{gasperich_h4,genovese_h4-agp}. Viewed this way, $H_2$ and $H_4$ are not merely small accuracy tests, but a minimal hierarchy of nodal difficulty: two electrons test whether the basis-free geminal and Jastrow are sufficiently flexible, whereas four electrons already expose when a compact pairing determinant no longer supplies the optimal nodal topology.

	\subsection{Hydrogen molecule dissociation}
		\label{ssec:h2_results}
	\begin{figure}
		\subfloat[]{\includegraphics[scale=0.525]{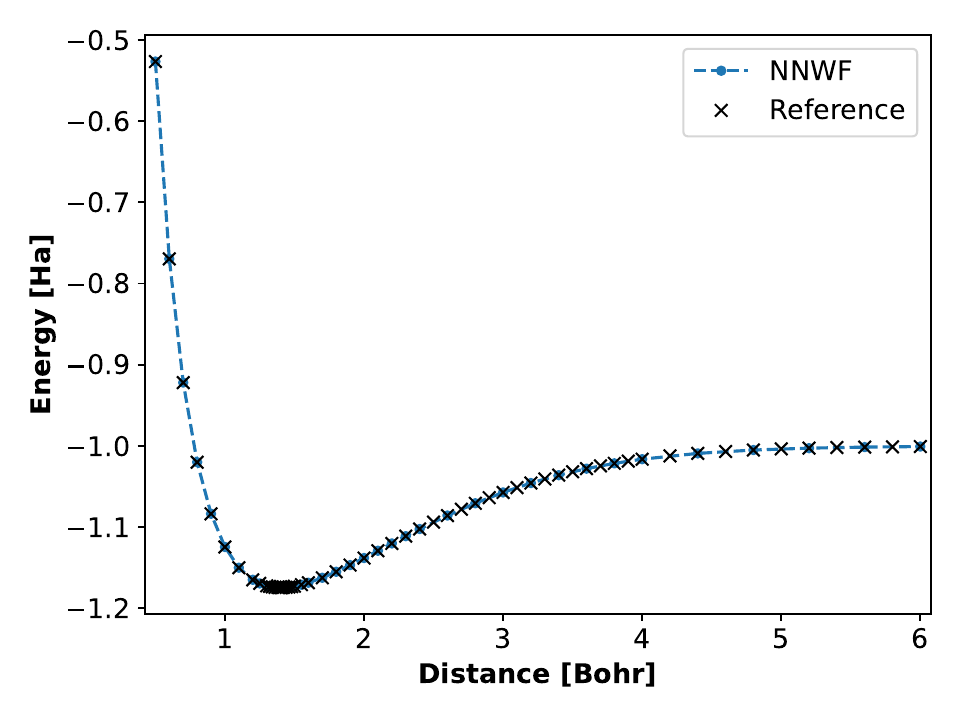} \label{subfig:h2_e_dx}} \hfill
		\subfloat[]{\includegraphics[scale=0.525]{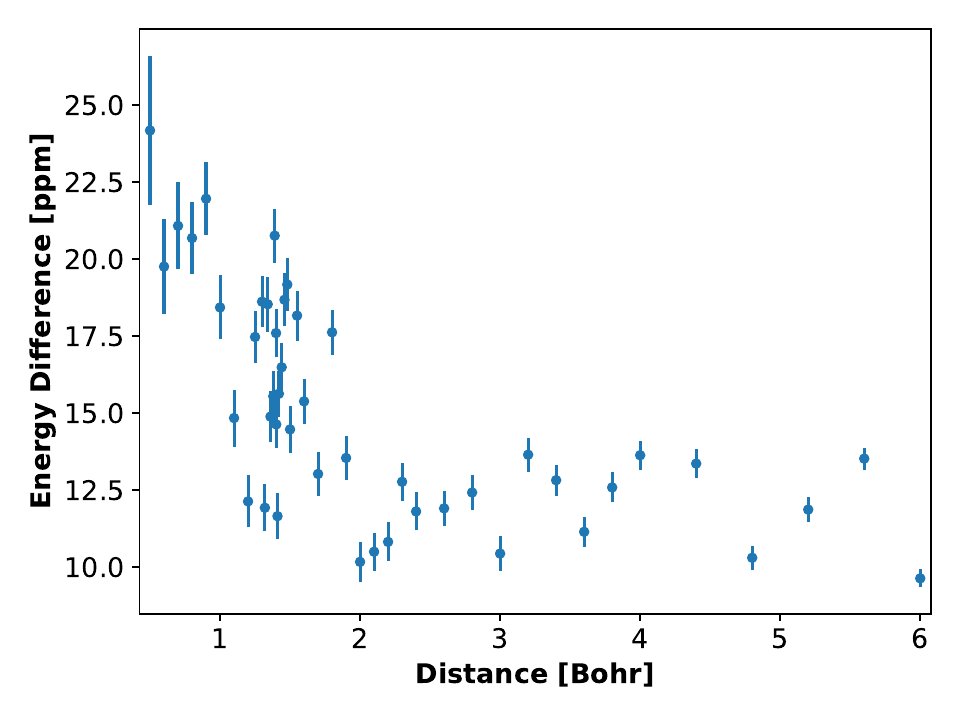} \label{subfig:h2_ediff_dx}}
		\caption{$H_2$: VMC energies (a) and relative energy differences (b) of NNAGP + all-body-NNJF for different H--H distances, compared with reference energies \cite{pachucki_born-oppenheimer_2010}. \label{fig:h2_e_dx}}
	\end{figure}
	As a first validation, we computed the $H_2$ dissociation curve using the two-particle NNAGP + all-body-NNJF ansatz with $80\times80$ hidden units. The energies are shown in Fig.~\ref{subfig:h2_e_dx} together with virtually exact reference values \cite{pachucki_born-oppenheimer_2010}. On this energy scale the VMC and reference curves are visually indistinguishable, showing that the NNWF captures nearly all correlation energy for this two-electron system.
	The equilibrium region is dominated by dynamical correlation on top of a single, well-defined covalent bond, whereas bond stretching gradually turns the problem into a static-correlation test because the covalent and ionic valence-bond structures become nearly degenerate, as also emphasized in our earlier fermionic shadow-wave-function study of stretched $H_2$ \cite{Calcavecchia2015ShadowSlater}.
	
	Figure~\ref{subfig:h2_ediff_dx} resolves the remaining deviations. The relative error increases below 2 Bohr, i.e., around and below the equilibrium distance of about 1.4 Bohr, reaching approximately $0.0025\%$ or $0.025$ mHa. At larger separations the error remains close to $0.001\%$. This behavior differs from many conventional single-reference electronic-structure methods, which are often accurate near equilibrium but deteriorate for stretched $H_2$ as static correlation becomes important \cite{foulkes_QMC}. For a two-electron singlet, however, this distinction is less severe than in larger systems: in a finite orbital basis, the coupled-cluster singles-and-doubles excitation manifold already spans the full-CI space \cite{BartlettMusial2007CC}, so $H_2$ is a useful validation problem but not the decisive multireference stress test. In the two-electron case, however, the AGP determinant in Eq.~\eqref{eq:psi_jagp} reduces to a single geminal $\Psi_G$. As discussed in our previous work \cite{kessler_2particle-nnvmc}, the resulting NNWF is an arbitrarily flexible approximator for exchange-symmetric two-particle ground states. The remaining error is therefore dominated by the short-range dynamical correlation near the equilibrium geometry rather than by a static-correlation failure. The dependence on network size is analyzed in Sec.~\ref{ssec:nn_size_scaling}.
	
	\subsection{Rectangular hydrogen tetramer}
	\label{ssec:h4_results}
	\begin{figure}
		\includegraphics[scale=0.48]{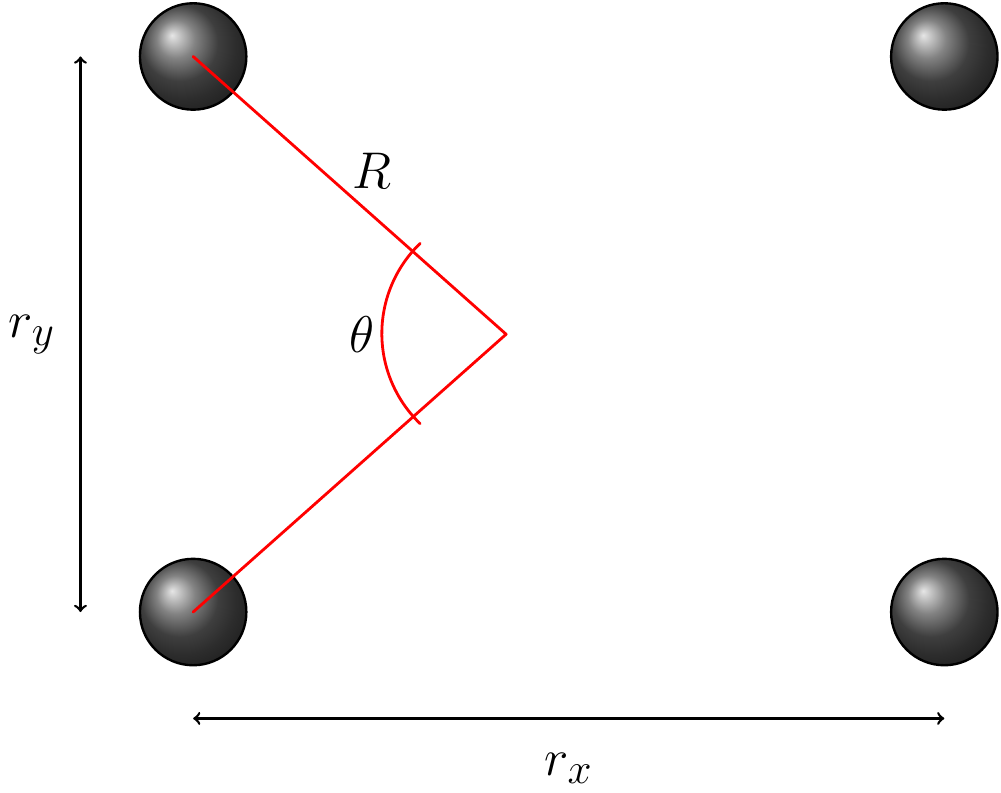}	
		\caption{Geometry and parametrizations of the rectangular $H_4$ benchmark system.\label{fig:h4_rect}}
	\end{figure}
	\begin{figure}
		\includegraphics[scale=0.53]{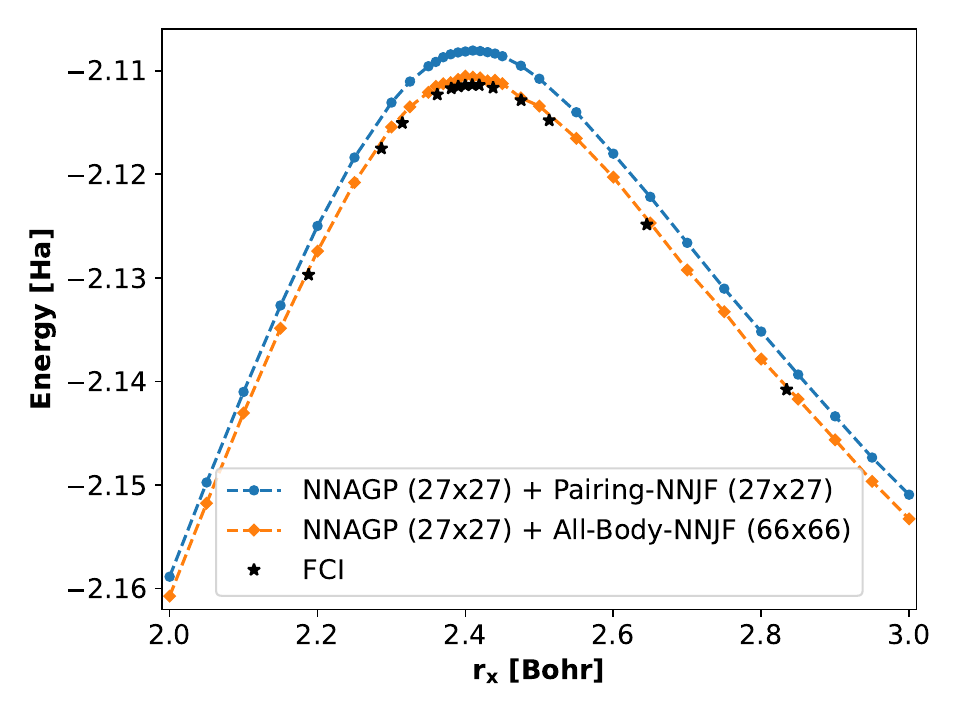}
		\caption{$H_4$: VMC energies of NNWFs for different H--H distances in the $x$ direction, with $r_y=2.4$ Bohr. FCI energies \cite{genovese_h4-agp} are shown as reference values. \label{fig:h4_e_dx}}
	\end{figure}
	\begin{figure}
		\includegraphics[scale=0.53]{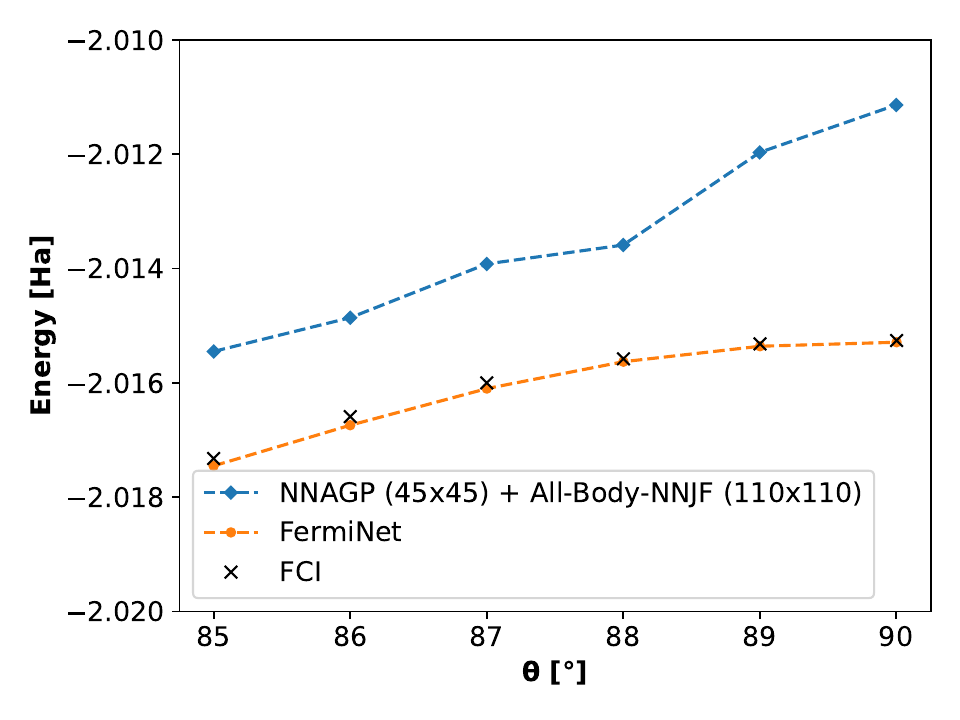}
		\caption{$H_4$: VMC energies of the NNWF for different H--H angles at fixed radius $R=3.2843$ Bohr. FermiNet and FCI energies from Ref.~\onlinecite{Pfau_FermiNet} are shown for comparison. \label{fig:h4_e_theta}}
	\end{figure}
	\begin{table*}
		\caption{\label{tab:h4_energies}Energies in Hartree for $H_4$ at fixed $r_y=2.4$ Bohr and varying $r_x$. VMC results from NNAGP+pairing-NNJF [NNWF(P)] and NNAGP+all-body-NNJF [NNWF(A)] are compared with conventional AGP+pairing-Jastrow results from Ref.~\onlinecite{genovese_h4-agp} at the VMC [JAGP(V)] and DMC [JAGP(D)] levels, both with a cc-pVDZ basis, and with full-CI values in a cc-pVQZ basis.}
		\begin{ruledtabular}
			\begin{tabular}{lccccc}
				$r_x$ & NNWF(P) & NNWF(A) & JAGP(V) & JAGP(D) & FCI \\
				2.188&-2.12685(1)&-2.12932(1)&            & -2.1307(1)& -2.1297\\
				2.4  &-2.10815(1)&-2.11052(1)&-2.1075(4)  & -2.1125(2)& -2.1114\\
				2.646&-2.12182(1)&-2.12477(1)&            & -2.1257(1)& -2.1248\\
				3.0  &-2.15093(1)&-2.15327(1)&-2.1491(3)  &            &\\
			\end{tabular}
		\end{ruledtabular}
	\end{table*}
	We next consider the rectangular $H_4$ system in the two parametrizations shown in Fig.~\ref{fig:h4_rect}. In the first case, the separation between the two pairs of atoms in the $y$ direction is fixed at $r_y=2.4$ Bohr, while the distance in the $x$ direction, $r_x$, is varied. Ref.~\onlinecite{genovese_h4-agp} used these configurations to test a conventional JAGP wave function with basis-set geminal and pairing terms. Table~\ref{tab:h4_energies} compares our VMC energies with their VMC, DMC, and FCI data, and Fig.~\ref{fig:h4_e_dx} shows the corresponding energy curves.
	This four-electron system is a more discriminating benchmark than $H_2$ because the relative importance of dynamical and static correlation can be varied geometrically. This tunability is precisely why rectangular $H_4$, or equivalently interacting $(H_2)_2$, has been used to assess fixed-node errors and configuration mixing in DMC calculations \cite{gasperich_h4} and the accuracy of the Jastrow--AGP form \cite{genovese_h4-agp}. When two H$_2$-like units remain internally well bound and are brought into weak contact, the shallow stabilization is mainly dynamical and dispersion-like, a type of weak binding for which QMC has proven particularly useful in molecular benchmarks \cite{Sorella2007VdWQMC}. By contrast, increasing the intramolecular H--H distance or approaching the large-radius square geometry increases static, multireference correlation by making several pairing patterns nearly degenerate.
	
	The all-body-NNJF gives consistently lower VMC energies than the pairing-NNJF. Since the NNAGP part is identical in both cases, this directly demonstrates that the all-body Jastrow factor recovers substantially more correlation energy. For the smaller $H_4$ rectangles, the all-body-NNJF nearly reaches the FCI reference, consistent with the flexibility expected from the universal approximation argument in Sec.~\ref{sec:nnagp}.
	The improvement should therefore be interpreted primarily as better recovery of dynamical correlation at a fixed nodal surface, including the dispersion-like contribution responsible for the shallow minimum of the compact rectangular scan in Fig.~\ref{fig:h4_e_dx}.
	
	Table~\ref{tab:h4_energies} also isolates the effect of replacing basis expansions by neural networks. At $r_x=2.4$ and $3.0$ Bohr, the NNWF with pairing-NNJF slightly improves over the structurally analogous basis-set JAGP VMC result. This confirms that neural-network components can improve the representational flexibility of a compact geminal ansatz, as previously observed for neural-network Slater--Jastrow wave functions \cite{Pfau_FermiNet}. Whether this is more efficient than increasing the conventional one-particle basis, for example to cc-pVTZ in Ref.~\onlinecite{genovese_h4-agp}, remains a separate optimization question.
	
	The all-body-NNJF energies are within 1 mHa of the FCI values and 1--2 mHa below the DMC energies obtained with the conventional JAGP ansatz. Thus, for these geometries, a VMC calculation with a sufficiently flexible positive Jastrow factor can approach the accuracy expected from a fixed-node projection method. This also indicates that the AGP determinant provides an accurate nodal surface for the considered $H_4$ rectangles; the dominant remaining control parameter is the all-body-NNJF size, discussed in Sec.~\ref{ssec:nn_size_scaling}.

	The second parametrization, used in Ref.~\onlinecite{Pfau_FermiNet} to validate FermiNet, keeps the four hydrogen atoms on a circle of radius $R=3.2843$ Bohr while varying the angle $\theta$ between neighboring atoms \cite{troy_gordon_h4_CCD}. These geometries are more stretched than those in the first benchmark. They therefore shift the dominant error source from dynamical correlation, which a positive Jastrow factor can recover efficiently, toward static correlation and nodal structure. Figure~\ref{fig:h4_e_theta} compares our NNWF with all-body-NNJF to the FermiNet and FCI data. For $\theta$ between about $85^\circ$ and $88^\circ$, the NNWF remains approximately 2 mHa above FermiNet and FCI, which are indistinguishable on this scale. The deviation increases further as the rectangle approaches the square geometry at $\theta=89^\circ$ and $90^\circ$.
	Because the all-body-NNJF already displayed DMC-like correlation recovery for the smaller rectangles, this remaining error is most naturally assigned to the AGP nodal surface. This is also the geometry where other methods can predict a nonphysical cusp in the energy curve due to a crossing of Hartree--Fock states \cite{burton_h4}, and, at large radius, even an incorrect energetic minimum \cite{troy_gordon_h4_CCD}. The AGP-based NNWF does not introduce such qualitative artifacts, but its accuracy is limited by insufficient nodal flexibility for the large-radius square.
	
	\subsection{Accuracy and cost versus neural-network size}
	\label{ssec:nn_size_scaling}
	\begin{figure*}
		\subfloat[]{\includegraphics[scale=0.54]{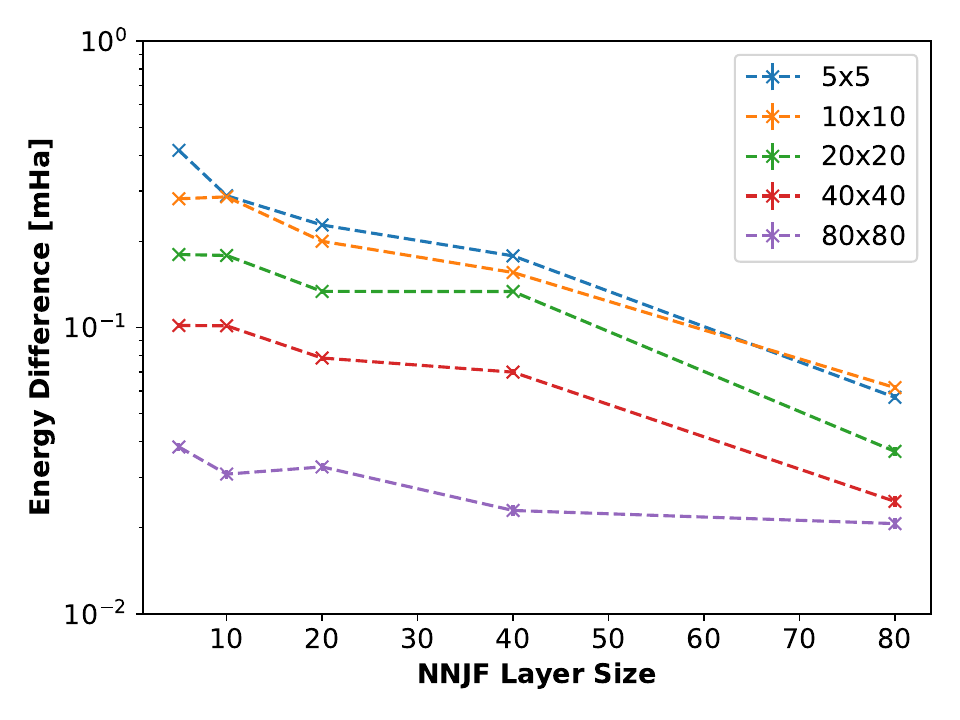}\label{subfig:h2_comp_nunits}} \hfill
		\subfloat[]{\includegraphics[scale=0.56]{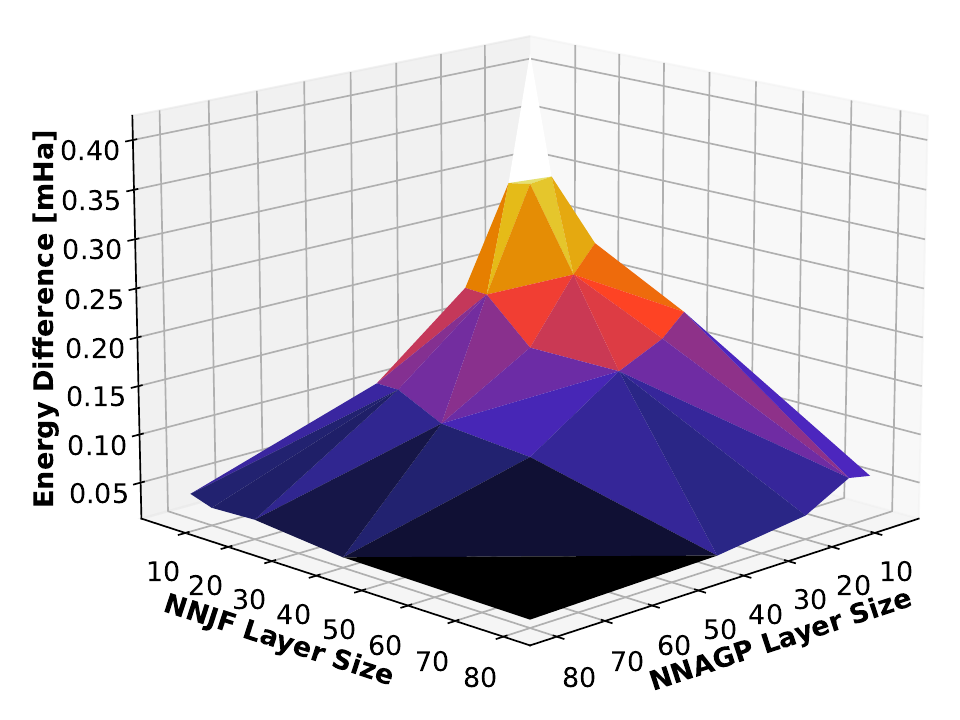}\label{subfig:h2_comp_nunits_3d}} \hfill
		\subfloat[]{\includegraphics[scale=0.54]{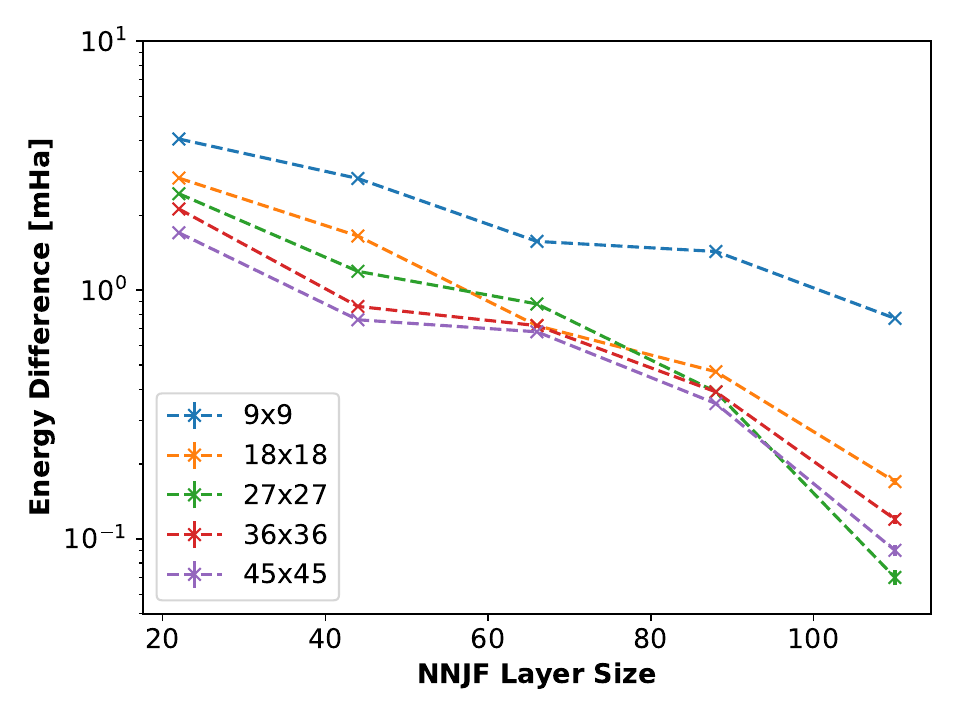}\label{subfig:h4_comp_nunits}} \hfill
		\subfloat[]{\includegraphics[scale=0.56]{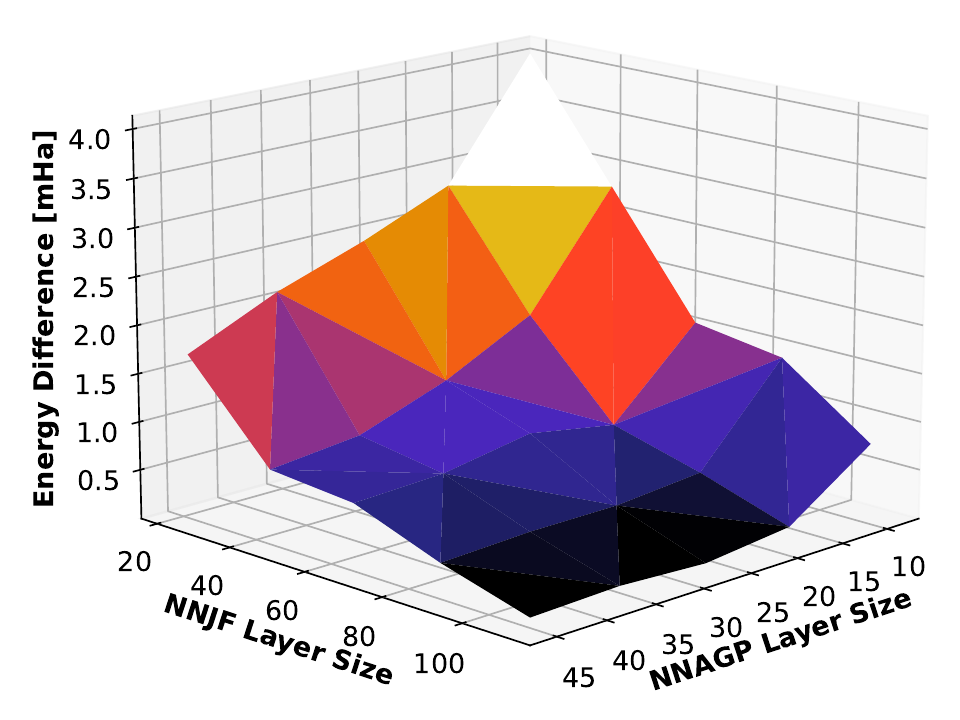}\label{subfig:h4_comp_nunits_3d}} 
		\caption{Energy differences to reference values \cite{pachucki_born-oppenheimer_2010,genovese_h4-agp} for NNAGP + all-body-NNJF wave functions with different numbers of hidden units. $N_G$ denotes the NNAGP size and $N_J$ the NNJF size. Panels (a) and (b) show $H_2$; panels (c) and (d) show $H_4$ at $r_x=r_y=2.4$ Bohr. \label{fig:h2_h4_comp_nunits}}
	\end{figure*}
	\begin{figure}[!htbp]
		\includegraphics[scale=0.45]{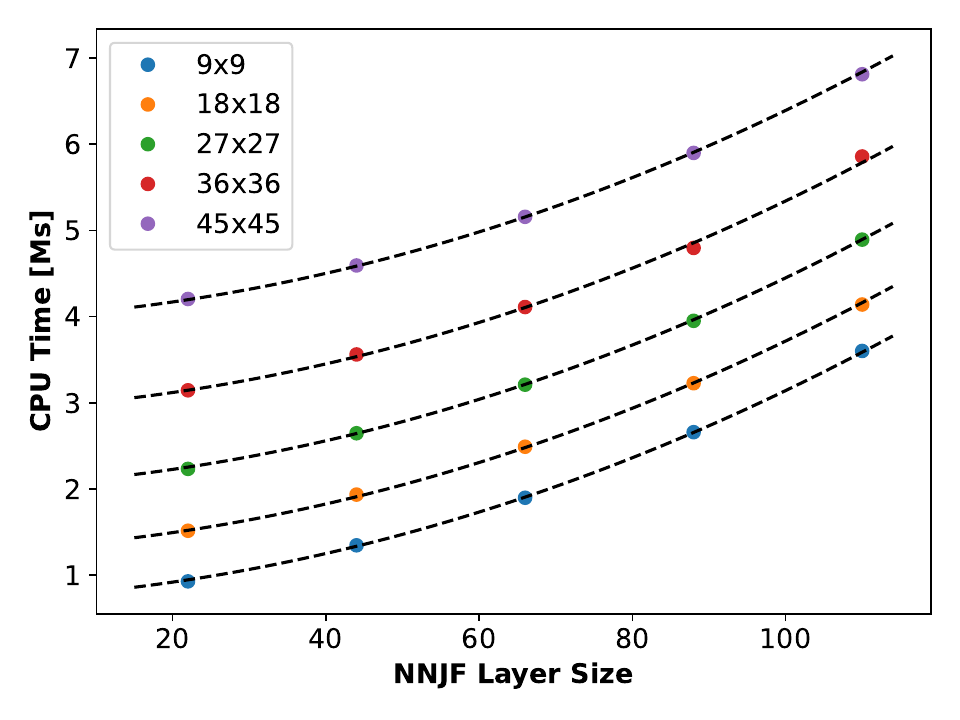}
		\caption{$H_4$ / NNAGP + all-body-NNJF: total CPU time for energy minimization as a function of the numbers of hidden units in the NNAGP and NNJF. Lines show a fitted second-order polynomial model.\label{fig:h4_cputime}}
	\end{figure}
	To disentangle the roles of the AGP and Jastrow parts, Fig.~\ref{fig:h2_h4_comp_nunits} shows the energy errors obtained with different numbers of hidden units in the NNAGP and NNJF networks.
	
	For $H_2$, the error decreases nonlinearly with the size of either the NNAGP or the NNJF, appearing approximately linear on the logarithmic scale. The NNAGP size has a slightly stronger effect, although the distinction between geminal and Jastrow flexibility is less sharp in a two-particle system.
	
	For square $H_4$ at $r_x=r_y=2.4$ Bohr, the three-dimensional plot shows a clearer separation. Increasing the NNJF size substantially improves the energy nearly independently of the NNAGP size, whereas increasing the NNAGP size saturates around $18\times18$ hidden units. This supports the interpretation of Sec.~\ref{ssec:h4_results}: for compact $H_4$ geometries, a single AGP determinant already gives a high-quality nodal surface, so that the positive NNJF can recover most of the remaining correlation energy. By contrast, even the largest networks considered here leave an error of $\gtrsim2$ mHa for the large-radius $H_4$ square, indicating that additional nodal flexibility, rather than a larger positive Jastrow factor alone, is required.
		
	Finally, Fig.~\ref{fig:h4_cputime} shows the total CPU time for wave-function optimization as a function of the numbers of hidden units, $N_G$ and $N_J$, in the NNAGP and NNJF. The number of optimization iterations was kept fixed, so that the timing mainly reflects network-evaluation cost. We fit the data to
	\begin{equation*}
		T(N_{G}, N_{J}) = T_0 + a_G N_{G} + b_G N_{G}^2 + a_J N_{J} + b_J N_{J}^2 ,
	\end{equation*}
	where the linear and quadratic terms reflect the growth in network connections. The fitted parameters are $T_0\approx0.32$ Ms, $a_G\approx0.037$ Ms, $b_G\approx0.00098$ Ms, $a_J\approx0.0054$ Ms, and $b_J\approx0.00019$ Ms. Adding a hidden unit to the NNAGP is therefore several times more expensive than adding one to the NNJF, as expected because the geminal is evaluated repeatedly in constructing the AGP determinant. Relating cost to accuracy suggests an approximately inverse relation between energy error and CPU time over the range studied, with larger NNJFs becoming increasingly efficient relative to smaller ones.
	 
	\section{Conclusions and outlook}
	We have introduced basis-free Jastrow--AGP many-body wave functions in which FFNNs replace conventional basis-set expansions in both the geminal and the Jastrow factor.
	The construction preserves the compact structure and interpretability of a single AGP determinant while adding systematically improvable neural-network correlation factors.
	In addition to reproducing the pairing-Jastrow structure of conventional JAGP wave functions, we introduced an all-body neural-network Jastrow factor that can recover correlation energy at fixed nodes within ordinary VMC. We applied the ansatz to $H_2$ and to rectangular $H_4$ benchmark geometries and compared the results with high-accuracy reference data.
	
	For $H_2$, the FFNN-based wave function remains highly accurate along the dissociation curve, including stretched configurations where static correlation is essential. With the chosen network sizes, the relative error remains between $0.001\%$ and $0.0025\%$, corresponding to an absolute deviation below $0.025$ mHa. The network-size study confirms systematic improvement with the number of hidden units.
	Because this is a two-electron problem, however, the absence of a stretched-bond failure should not be overinterpreted: the AGP geminal already has enough nodal flexibility, and the remaining error mostly tests how efficiently the neural Jastrow factor represents dynamical correlation.
	
	For $H_4$, replacing the basis-set geminal and pairing functions by neural networks improves or matches the corresponding conventional JAGP VMC energies. The all-body-NNJF further lowers the energy and reaches FCI-quality results for compact rectangular geometries, demonstrating that a positive neural-network Jastrow factor can recover most of the missing correlation energy when the AGP nodal surface is adequate. The scaling analysis shows that, in this regime, increasing the NNJF size is more important than further enlarging the NNAGP.
	The hydrogen tetramer is therefore the more informative diagnostic: by changing the intra- and interfragment H--H distances, it exposes when the calculation is limited by dynamical correlation recovery and when it is limited by static, multireference correlation encoded in the nodal surface.
	
	The large-radius $H_4$ square reveals the limitation of the present compact ansatz. Even with the largest networks used here, the energy remains more than 2 mHa above the FermiNet and FCI references. Since the all-body-NNJF is already sufficiently flexible to recover DMC-like correlation for compact rectangles, this residual error points to an AGP nodal limitation. Thus, the method succeeds as a compact and efficient neural JAGP ansatz, while also identifying where additional determinant, backflow, or other nodal degrees of freedom would be needed.
	In this respect, the present basis-free construction should be viewed as complementary to the TurboRVB/JAGP and shadow-wave-function developments: it retains the transparent AGP--Jastrow separation, but replaces fixed analytical or basis-expanded building blocks by neural-network representations.
	The main conceptual outcome is therefore a diagnostic one: neural-network flexibility is most useful when it can be assigned to a specific physical role. In the present ansatz, the Jastrow network mostly measures recoverable dynamical correlation at fixed nodes, whereas the AGP network and any future nodal extensions determine how far the ansatz can be pushed into genuinely multireference regimes.
	
	Overall, the combination of AGP determinants, neural-network Jastrow factors, and VMC provides a promising route toward accurate and efficient simulations with compact wave functions. Future improvements should focus on optimized GPU evaluation, more advanced stochastic optimization, and controlled extensions of the nodal surface while retaining the transparent separation between nodal structure and positive correlation factor. For extended or weakly interacting systems, an equally important direction is to build size consistency into the ansatz, for example through localized or block-separable geminals and additive logarithmic Jastrow architectures that satisfy the correct cluster decomposition by construction.
	
	\begin{acknowledgments}
	The authors gratefully acknowledge the computing time made available to them on the high-performance computer [Cluster name/s] at the NHR Center Paderborn Center for Parallel Computing (PC2). This center is jointly supported by the Federal Ministry of Research, Technology and Space and the state governments participating in the National High-Performance Computing (NHR) joint funding program (\url{www.nhr-verein.de/en/our-partners}).
	\end{acknowledgments}

    \bibliography{references_nnvmc_jagp_pre}

\end{document}